\documentclass[reprint,amsmath,amssymb,aps,pre,longbibliography]{revtex4-1}

\usepackage[dvips]{graphicx}
\usepackage{color}
\usepackage{pstricks,pst-grad}
\usepackage{amsbsy}
\usepackage{epic}
\usepackage[normalem]{ulem}
\usepackage{bm}

\usepackage{hyperref}
\usepackage[hyphenbreaks]{breakurl}

\newcommand\beq{\begin{equation}}
\newcommand\eeq{\end{equation}}
\newcommand\beqa{\begin{eqnarray}}
\newcommand\eeqa{\end{eqnarray}}
\newcommand{\nn}{\nonumber\\}
\newcommand{\dd}{\text{d}}
\newcommand{\II}{\mathcal{I}}
\def\bal#1\eal{\begin{align}#1\end{align}}

\newcommand{\vvss}{\vspace{-3mm}\\}

\begin{document}

\title{Finite-size estimates of Kirkwood-Buff and similar integrals}

\author{Andr\'es Santos}
\email{andres@unex.es}
\homepage{http://www.unex.es/eweb/fisteor/andres/}
\affiliation{Departamento de F\'{\i}sica  and Instituto de Computaci\'on Cient\'{\i}fica Avanzada (ICCAEx), Universidad de
Extremadura, Badajoz, E-06006, Spain}

\begin{abstract}
Recently, Kr\"uger and  Vlugt [Phys.\ Rev.\ E \textbf{97}, 051301(R) (2018)] have proposed a method  to approximate an improper integral $\int_0^\infty \text{d}r\, F(r)$, where $F(r)$ is a given  oscillatory function, by a finite-range integral $\int_0^L \text{d}r\, F(r) W(r/L)$ with an appropriate weight function $W(x)$. The method is extended here to an arbitrary (embedding) dimensionality $d$. A study of three-dimensional Kirkwood-Buff integrals, where $F(r)=4\pi r^2h(r)$, and static structure factors, where $F(r)=(4\pi/q) r\sin(qr) h(r)$, $h(r)$ being the  pair correlation function, shows that, in general, a choice $d\neq 3$ (e.g., $d=7$) for the embedding dimensionality may significantly reduce the error of the approximation $\int_0^\infty \text{d}r\, F(r)\simeq \int_0^L \text{d}r\, F(r) W(r/L)$.

\end{abstract}

\date{\today}


\maketitle

\section{Introduction}
In the statistical-mechanical theory of liquids, the Kirkwood-Buff (KB) integral plays a distinguished role \cite{KB51,BN06,BNS09}. It has the form
\begin{equation}
\label{1}
\II[F(r)]\equiv \int_0^\infty \dd r\,F(r),
\end{equation}
where, in the case of three-dimensional (3D) systems, $F(r)=4\pi r^2 h(r)$, $h(r)$ being the (total) pair correlation function \cite{BH76,HM06,S16}.
The KB integral is the zero-wave-number limit ($q\to 0$) of the Fourier transform of $h(r)$, defined as
\begin{align}
\label{hq}
\widetilde{h}(q)=&\int \dd^d\mathbf{r}\, e^{-i\mathbf{q}\cdot\mathbf{r}}h(r)\nn
=&(2\pi)^{d/2}\int_0^\infty \dd r\, r^{d-1}\frac{J_{d/2-1}(qr)}{(q r)^{d/2-1}}h(r),
\end{align}
where $d$ is the number of spatial dimensions and $J_\nu(x)$ is the Bessel function of the first kind. Therefore, $\widetilde{h}(q)$ has the structure of Eq.\ \eqref{1}, this time (again for 3D systems, $d=3$) with $F(r)=({4\pi}/{q})r\sin(qr)h(r)$. Apart from the limit $q\to 0$, the physical importance of $\widetilde{h}(q)$ lies in its direct relation to the static structure factor \cite{HM06,S16}, namely
\begin{equation}
\label{Sq}
S(q)=1+\rho \widetilde{h}(q),
\end{equation}
where $\rho$ is the number density of the fluid.

If $h(r)$ is obtained from computer simulations or from numerical solutions of integral-equation theories, its knowledge is limited to a finite range $r<L$, so that the conventional method consists in estimating the KB integral or the structure factor by a truncated integral, i.e.,
\begin{equation}
\label{0}
\II[F(r)]\simeq  \int_0^L \dd r\,F(r).
\end{equation}
On the other hand, the correlation function $h(r)$ is usually \emph{oscillatory}, which generally makes the convergence of the estimate \eqref{0} rather slow. It is then highly desirable to devise alternative approximate methods to estimate improper integrals of the form $\II[F(r)]$ that, while relying upon the knowledge of $F(r)$ for $r<L$ only, are much more efficient than Eq.\ \eqref{0}. The general problem of computing highly oscillatory integrals has aroused a large body of work by applied mathematicians, as summed up in a recent monograph \cite{DHI17}.
A method recently proposed in physics literature \cite{KSBKVS13,KV18} consists in approximating Eq.\ \eqref{1} by a finite-size integral of the form
\begin{equation}
\label{2}
\II_L[F(r)]\equiv \int_0^L \dd r\,F(r) W(r/L),
\end{equation}
with an appropriate \emph{weight} function $W(x)\neq 1$.

Of course, the computational problem described above is not limited to KB integrals and structure factors but extends, with different physical interpretations of the isotropic oscillatory function $F(r)$,  to other branches of physics where improper integrals of the form \eqref{1} are relevant. In those other more general cases, $r$ could not represent a spatial variable but, for instance, a wave number or a time variable.

In Ref.\ \cite{KV18}, Kr\"uger and Vlugt proposed a simple, practical, and accurate general prescription to approximate an improper integral of the form \eqref{1} by the finite-size integral \eqref{2}, where the weight function $W(x)$ is given by
\begin{equation}
\label{3}
W_3^{(2)}(x)=1 - \frac{23x^3}{8} + \frac{3 x^4}{4} + \frac{9 x^5}{8}.
\end{equation}
More specifically,
\begin{equation}
\label{3b}
\II[F(r)]= \II_L[F(r)]+{O}(L^{-3}).
\end{equation}

Let me rephrase and summarize the two main steps leading to the derivation of Eqs.\ \eqref{3} and \eqref{3b}. First, it is tacitly assumed that $\II[F(r)]$ comes from the 3D volume integral
\begin{equation}
\label{4}
\II[F(r)]=\int \frac{\dd^3\mathbf{r}}{4\pi r^2} {F(r)},
\end{equation}
after passing to spherical coordinates and integrating over the angular variables. Next, use is made of the authors' proof  that
\begin{equation}
\label{5}
\int \frac{\dd^3\mathbf{r}}{4\pi r^2}{F(r)}y_3(r/L)=\II[F(r)]-\frac{3}{2L}\II[F(r)r]+{O}(L^{-3}),
\end{equation}
where $4\pi y_3(x)$ is the intersection volume of two spheres of unit diameter separated a distance $x$, i.e.,
\begin{equation}
\label{6}
y_3(x)=\left(1-\frac{3x}{2}+\frac{x^3}{2}\right)\Theta(1-x).
\end{equation}
Actually, the proof in Ref.\ \cite{KV18} extends Eq.\ \eqref{5} to non-spherical shapes, in which case the function $y_3(x)$ depends on the particular shape, $L=6V/A$ ($V$ and $A$ being the volume and surface area, respectively), and, in general, ${O}(L^{-3})\to {O}(L^{-2})$. On the other hand, a spherical shape, and hence Eq.\ \eqref{6}, is needed for the derivation of Eq.\ \eqref{3} as
\begin{equation}
\label{7}
W_3^{(2)}(x)=y_3(x)\left(1+\frac{3x}{2}+\frac{9x^2}{4}\right).
\end{equation}

\begin{table}[htb]
\caption{Coefficient $a_d$ and function $y_d(x)$ for the first few odd values of $d$. The Heaviside function $\Theta(1-x)$ is omitted for clarity.}
\label{table:1}
\begin{ruledtabular}
\begin{tabular}{lll}
$d$ & $a_d$&$y_d(x)$\\
\hline
$1$&$1$&$1-x$\\
\vvss
$2$&$\displaystyle{\frac{4}{\pi}}$&$\displaystyle{\frac{2}{\pi}\left(\cos^{-1}x-{x}\sqrt{1-x^2}\right)}$\\
\vvss
$3$&$\displaystyle{\frac{3}{2}}$&$\displaystyle{(1-x)^{2}\left(1+\frac{x}{2}\right)}$\\
\vvss
$4$&$\displaystyle{\frac{16}{3\pi}}$&$\displaystyle{\frac{2}{\pi}\left[\cos^{-1}x-\frac{x}{3}\sqrt{1-x^2}(5-2x^2)\right]}$\\
\vvss
$5$&$\displaystyle{\frac{15}{8}}$&$\displaystyle{(1-x)^{3}\left(1+\frac{9x}{8}+\frac{3x^2}{8}\right)}$\\
\vvss
$6$&$\displaystyle{\frac{32}{5\pi}}$&$\displaystyle{\frac{2}{\pi}\left[\cos^{-1}x-\frac{x}{15}\sqrt{1-x^2}(33-26x^2+8x^4)\right]}$\\
\vvss
$7$&$\displaystyle{\frac{35}{16}}$&$ \displaystyle{(1-x)^{4}\left(1 + \frac{29 x}{16} + \frac{5 x^2}{4} + \frac{5 x^3}{16}\right)}$\\
\vvss
$8$&$\displaystyle{\frac{256}{35\pi}}$&$\displaystyle{\frac{2}{\pi}\Big[\cos^{-1}x-\frac{x}{105}\sqrt{1-x^2}(279-326x^2}$\\
&&$\displaystyle{+200x^4-48x^6)\Big]}$\\
\vvss
$9$&$\displaystyle{\frac{315}{128}}$&$ \displaystyle{(1-x)^{5}\left(1 + \frac{325 x}{128} + \frac{345 x^2}{128} + \frac{175 x^3}{128}+ \frac{35 x^4}{128}\right)}$\\
\end{tabular}
\end{ruledtabular}
\end{table}

\begin{figure}[htb]
\includegraphics[width=8cm]{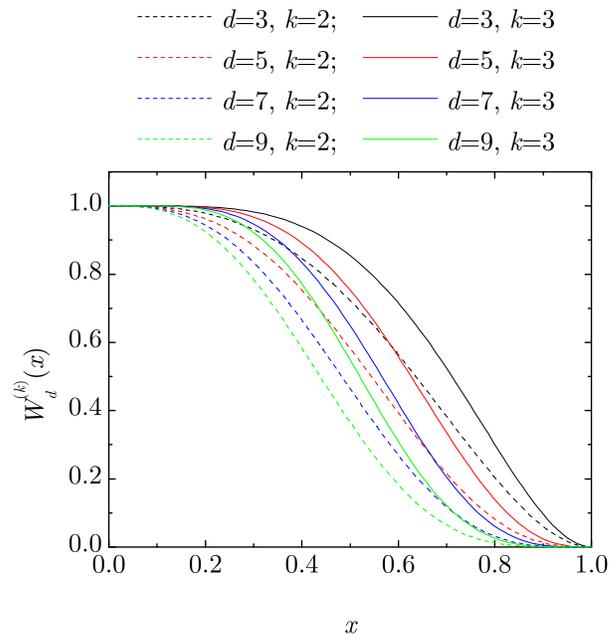}
\caption{Plot of the weight functions $W_d^{(k)}(x)$ for embedding dimensionalities $d=3$, $5$, $7$, and $9$ and indices $k=2$ and $3$.\label{fig:W}}
\end{figure}

In their paper \cite{KV18}, Kr\"uger and Vlugt motivate the result posed by Eqs.\ \eqref{3} and \eqref{3b} as a useful way to estimate 3D KB integrals, in which case $F(r)=4\pi r^2 h(r)$. On the other hand, as said before, the result is not restricted \emph{a priori} to 3D KB integrals, i.e., $F(r)$ can be in principle any function such that the (formally one-dimensional) integral $\II[F(r)]$ converges.
It is then tempting to wonder how the procedure summarized above would be generalized by freely assuming that the function $F(r)$ is embedded in a $d$-dimensional space and  rewriting $\II[F(r)]$ as a $d$-dimensional volume integral. The main goal of this paper is to perform such an extension and, additionally, show that a choice $d\neq 3$ allows one to obtain alternative weight functions $W(x)$   that are generally more efficient than Eq.\ \eqref{3}, even in the case of 3D KB integrals and structure factors.

The organization of the remainder of this paper is as follows. Section \ref{sec2} presents the extension to a generic dimensionality of the scheme devised in Ref.\ \cite{KV18}. This is followed in Sec.\ \ref{sec3} by a discussion on the application of the generalized method to the numerical or computational evaluation of KB integrals and static structure factors. Finally, the main conclusions of the paper are summarized in Sec.\ \ref{sec4}.

\section{Embedding in a $d$-dimensional space}
\label{sec2}
Let us assume that the isotropic function $F(r)$ is embedded in a vector space of $d$ dimensions.
In such a case, the counterpart of Eq.\ \eqref{4} is
\begin{equation}
\label{8}
\II[F(r)]=\int \frac{\dd^d\mathbf{r}}{\Omega_d r^{d-1}}F(r),
\end{equation}
where
$\Omega_d={2\pi^{d/2}}/{\Gamma(d/2)}$
is the total solid angle in $d$ dimensions.
Following essentially the same steps as done in Ref.\ \cite{KV18} to derive Eq.\ \eqref{5}, it is possible to generalize it as
\begin{equation}
\label{10}
\int \frac{\dd^d\mathbf{r}}{\Omega_d r^{d-1}}{F(r)}y_d(r/L)=\II[F(r)]-\frac{a_d}{L}\II[F(r)r]+{O}(L^{-3}),
\end{equation}
where
\begin{equation}
\label{11}
a_d\equiv\frac{2\pi^{-1/2}\Gamma(1+d/2)}{\Gamma(1/2+d/2)}
\end{equation}
and
$\Omega_d y_d(x)$ is the intersection volume of two $d$-dimensional spheres of unit diameter separated a distance $x$, so that $y_d(0)=1$ and $y_d(x)=0$ if $x\geq 1$. This quantity appears, for instance, in the context of the virial expansion of the pair correlation function \cite{S16}. Three equivalent representations of $y_d(r)$ are
\begin{align}
\label{yd}
y_d(x)=&I_{1-x^2}\left(1/2+d/2,{1}/{2}\right)\nn
&=1-a_d\int_0^x \dd t\,(1-t^2)^{(d-1)/2}\nn
&=a_d\int_x^1 \dd t\,(1-t^2)^{(d-1)/2},
\end{align}
where $I_z(a,b)=B_z(a,b)/B(a,b)$  is the {regularized}  incomplete beta function \cite{AS72,OLBC10}. If $d=\text{odd}$, $y_d(x)-1$ is an odd polynomial of degree $d$ \cite{BC87,TS03,TS06}, namely
\begin{equation}
\label{12}
y_d(x)=1-a_d\sum_{j=0}^{(d-1)/2}c_{j,d} x^{2j+1},
\end{equation}
where
\begin{equation}
c_{j,d}\equiv \frac{(-1)^j\Gamma(1/2+d/2)}{(2j+1)j!\Gamma(1/2+d/2-j)}.
\end{equation}
If $d=\text{even}$, Eq.\ \eqref{12}, with the upper summation limit $(d-1)/2$ replaced by $\infty$, gives the power series expansion of $y_d(x)$. In such a case ($d=\text{even}$), $y_d(x)$ can be more conveniently expressed as
\begin{equation}
y_d(x)=\frac{2}{\pi}\left[\cos^{-1}x-{x}\sqrt{1-x^2}P_{d/2-1}(x^2)\right],
\end{equation}
where
\begin{equation}
P_m(z)=\sum_{j=0}^m p_{j,m} z^j
\end{equation}
is a polynomial of degree $m$ with coefficients given by
$p_{0,m}=\frac{\pi}{2}a_{2m+2}-1$ and the recurrence relation ($j\geq 1$)
\begin{equation}
p_{j,m}=\frac{1}{2j+1}\left[2jp_{j-1,m}+\frac{\pi}{2}\frac{a_{2m+2}(-1)^j(m+1)!}{j!(m+1-j)!}\right].
\end{equation}
Regardless of whether $d$ is even or odd, it can be seen from Eq.\ \eqref{yd} that $y_d(x)\sim (1-x)^{(d+1)/2}$ in the region $x\lesssim 1$.

The explicit values of $a_d$ and expressions of $y_d(x)$ for embedding dimensionalities $1\leq d\leq 9$ are given in Table \ref{table:1}.
Henceforth, for simplicity, only the cases  $d=\text{odd}$ will be explicitly shown because the corresponding functions $y_d(x)$ are just polynomials. However, it can be checked that the results for $d=\text{even}$ follow patterns similar to those for $d=\text{odd}$. In fact,  the results obtained with dimensionalities $d$ and $d+1$ become closer and closer as $d$ increases.

The replacement $F(r)\to F(r) r^n$ in Eq.\ \eqref{10} yields (provided the integrals exist)
\begin{align}
\label{13}
\II[F(r)r^n]=&\int \frac{\dd^d\mathbf{r}}{\Omega_d r^{d-1}}F(r)r^ny_d(r/L)\nn
&+\frac{a_d}{L} \II[F(r)r^{n+1}]+{O}(L^{-3}).
\end{align}
Recursive application of Eq.\ \eqref{13} in Eq.\ \eqref{10} up to $n=k$ gives
\begin{equation}
\label{14}
\II[F(r)]= \II_{L,d}^{(k)}[F(r)]+{O}(L^{-3}),
\end{equation}
where
\begin{equation}
\label{15}
\II_{L,d}^{(k)}[F(r)]\equiv \int_0^L \dd r\,F(r) W_{d}^{(k)}(r/L),
\end{equation}
\begin{equation}
\label{16}
W_{d}^{(k)}(x)\equiv y_d(x)\sum_{n=0}^k (a_dx)^n.
\end{equation}
Equations \eqref{14}, \eqref{15}, and \eqref{16} generalize Eqs.\ \eqref{3b}, \eqref{2}, and \eqref{3}, respectively, which correspond to the particular choices $d=3$ and $k=2$. Incidentally,  the choice $d=1$ with $k=2$ leads to $W_1^{(2)}(x)=1-x^3$, which is the weight function proposed in Ref.\ \cite{KSBKVS13} by a different method.

Figure \ref{fig:W} shows the weight functions \eqref{16} with $d=3$, $5$, $7$, and $9$, and  $k=2$ and $3$. All of them have a similar qualitative shape, but,  due to the behavior $y_d(x)\sim (1-x)^{(d+1)/2}$, the curves have a flatter shape near $x=1$ as $d$ increases. While the functions $W_d^{(2)}(x)$ decay monotonically with increasing $x$, $W_d^{(3)}(x)$ present a practically unobservable maximum at $0.09<x<0.10$. This maximum, however, becomes more noticeable as $k$ increases (not shown). In fact, $W_d^{(k)}(x)\to y_d(x)(1-a_d x)^{-1}$ in the limit $k\to\infty$, so that it artificially diverges at a value $x<a_d^{-1}<1$. From a more practical of view, it turns out that the results obtained with $k\geq 4$ are generally worse than those obtained with $k=3$ (not shown).
Because of this, in what follows only the cases $k=2$ and $k=3$ will be explicitly considered.

Notice that
\begin{equation}
W_d^{(k)}(x)=1+{O}(x^{-3}),\quad k\geq 2,
\end{equation}
so that the influence of the choice of $d$ and $k\geq 2$ on $W_d^{(k)}(r/L)$ is of ${O}(L^{-3})$, i.e., of the same order as the terms neglected in Eq.\ \eqref{14}. On the other hand, from a practical point of view, the  error $\left|\II_{L,d}^{(k)}[F(r)]-\II[F(r)]\right|$ can be minimized by an appropriate choice of the embedding dimensionality $d$ and of the index $k$ for a given function $F(r)$ and a given cutoff distance $L$.

\begin{figure}[htb]
\includegraphics[width=8cm]{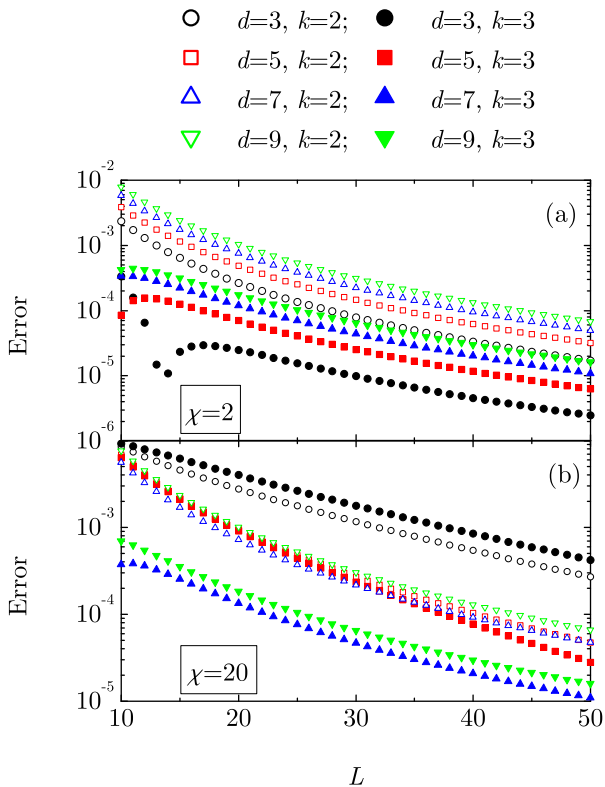}
\caption{Plot of the relative error $\left|\II_{L,d}^{(k)}[F(r)]/\II[F(r)]-1\right|$ versus $L$, where $F(r)=4\pi r^2 h(r)$ and $h(r)$ is given by Eq.\  \eqref{h(r)}. Panels (a) and (b) correspond to $\chi=2$ and $\chi=20$, respectively. Only the values corresponding to $L=\text{integer}$ are shown.\label{fig:Error_vs_L}}
\end{figure}

\begin{figure}[htb]
\includegraphics[width=8cm]{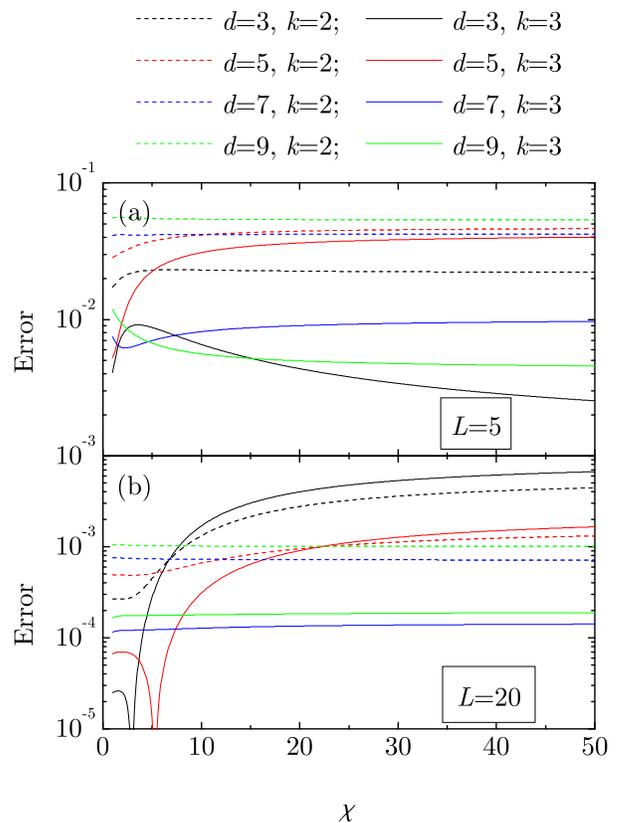}
\caption{Plot of the relative error $\left|\II_{L,d}^{(k)}[F(r)]/\II[F(r)]-1\right|$ versus $\chi$, where $F(r)=4\pi r^2 h(r)$ and $h(r)$ is given by Eq.\  \eqref{h(r)}. Panels (a) and (b) correspond to $L=5$ and $L=20$, respectively. \label{fig:Error_vs_chi} }
\end{figure}

\begin{figure}[htb]
\includegraphics[width=8cm]{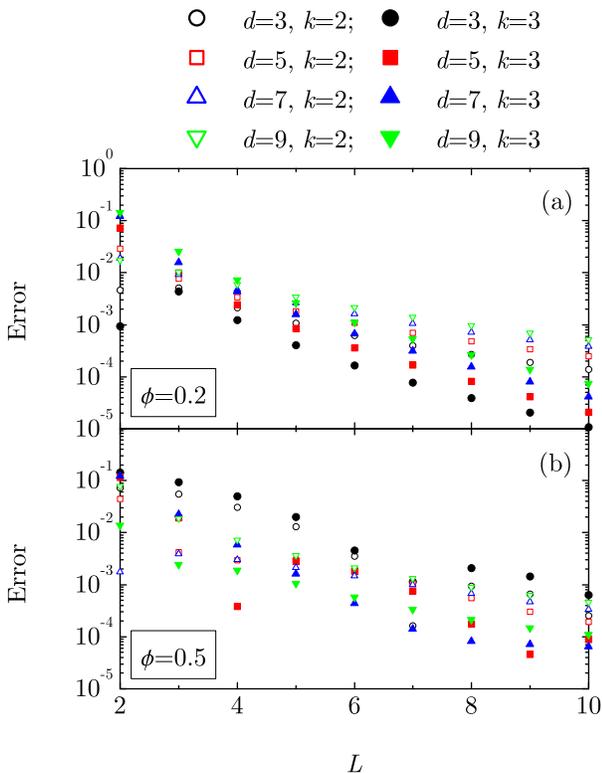}
\caption{Plot of the relative error $\left|\II_{L,d}^{(k)}[F(r)]/\II[F(r)]-1\right|$ versus $L$, where $F(r)=4\pi r^2 h(r)$ and $h(r)$ is the exact solution of the PY integral equation for hard spheres. Panels (a) and (b) correspond to $\phi=0.2$ and $\phi=0.5$, respectively. Only the values corresponding to $L=\text{integer}$ are shown.\label{fig:Error_vs_L_PY}}
\end{figure}

\begin{figure}[htb]
\includegraphics[width=8cm]{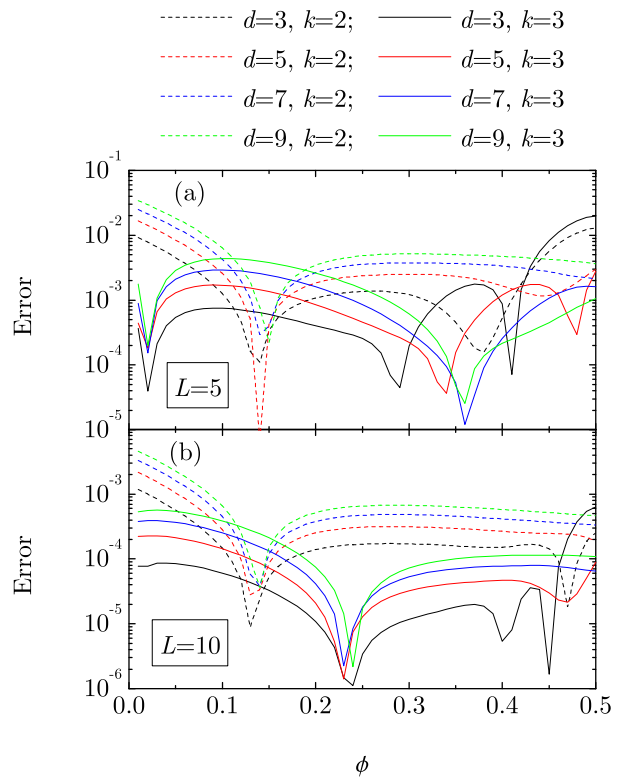}
\caption{Plot of the relative error $\left|\II_{L,d}^{(k)}[F(r)]/\II[F(r)]-1\right|$ versus $\phi$, where $F(r)=4\pi r^2 h(r)$ and $h(r)$ is the exact solution of the PY integral equation for hard spheres. Panels (a) and (b) correspond to $L=5$ and $L=10$, respectively. \label{fig:Error_vs_eta_PY}}
\end{figure}

\begin{figure}[htb]
\includegraphics[width=8cm]{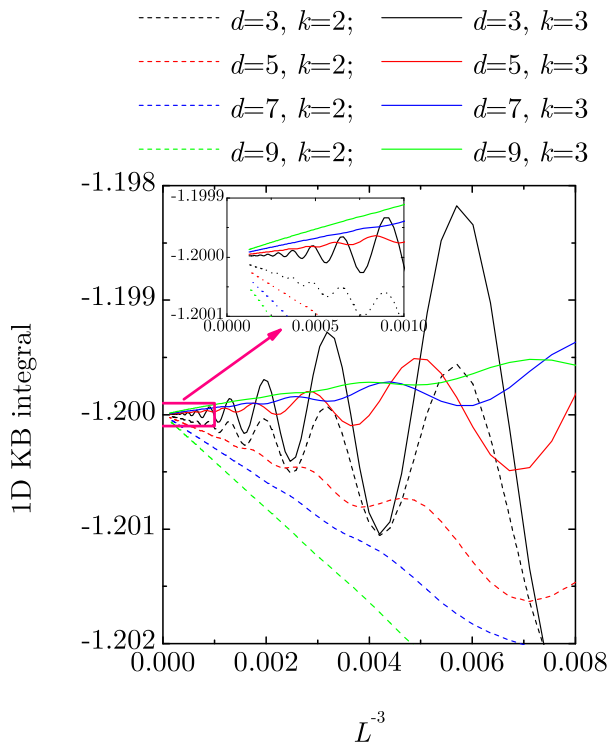}
\caption{Plot of the  $\II_{L,d}^{(k)}[F(r)]$ versus $L^{-3}$ ($L\geq 5$), where $F(r)=2 h(r)$ and $h(r)$ is the exact pair correlation function for a 1D system of hard rods at a packing fraction $\phi=0.8$. The inset is a magnification of the small framed region ($L\geq 10$) in the main figure. \label{fig:1D}}
\end{figure}

\begin{figure}[htb]
\includegraphics[width=8cm]{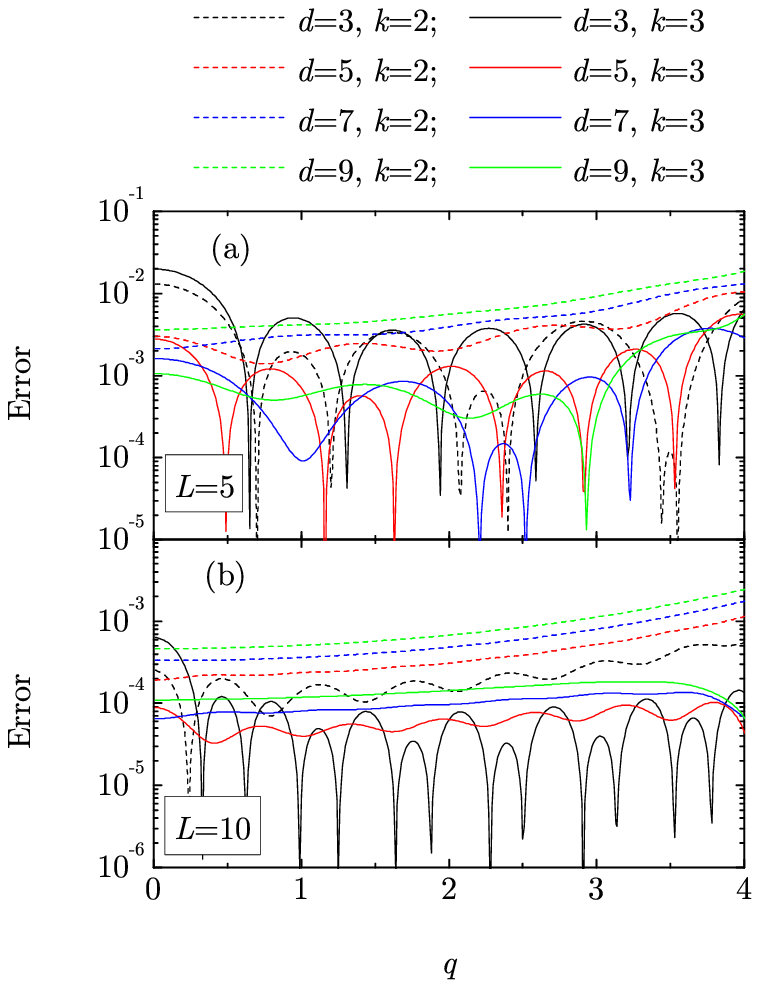}
\caption{Plot of the relative error $\left|\II_{L,d}^{(k)}[F(r)]/\II[F(r)]-1\right|$ versus $q$, where $F(r)=(4\pi/q) r\sin(qr) h(r)$ and $h(r)$ is the exact solution of the PY integral equation for hard spheres at a packing fraction $\phi=0.5$. Panels (a) and (b) correspond to $L=5$ and $L=10$, respectively. \label{fig:Error_vs_q_PY}}
\end{figure}

\section{Discussion}
\label{sec3}

One might reasonably argue that the choice of the embedding dimensionality $d$ in the approximation \eqref{14} must be dictated by the dimensionality of the physical problem underlying the evaluation of the (one-dimensional) integral $\II[F(r)]$. According to this reasoning, if the physical problem consists in the computation of the 3D KB integral, i.e., $F(r)=4\pi r^2 h(r)$, or of the 3D structure factor, i.e., $F(r)=(4\pi/q) r\sin(qr) h(r)$, then one should take $d=3$. On the other hand, from a strict mathematical point of view, the integral one wants to approximate by application of Eq.\ \eqref{14} is blind to the physical origin of the problem, so one can always assume that $F(r)$ is embedded in a higher-dimensional space.

\subsection{Three-dimensional Kirkwood-Buff integrals}

To further elaborate on the previous point, let us take $F(r)=4\pi r^2 h(r)$ and  consider the same model 3D pair correlation function  as given by Eq.\ (25) of Ref.\ \cite{KV18}, namely
\begin{equation}
\label{h(r)}
h(r)=
\begin{cases}
-1,&\displaystyle{r<\frac{19}{20}},\\
\displaystyle{\frac{3\cos\left[2\pi\left(r-\frac{21}{20}\right)\right]}{2r}e^{-(r-1)/\chi}},&r>\displaystyle{\frac{19}{20}},
\end{cases}
\end{equation}
where $\chi$ represents the correlation length. As said before, the discussion is restricted to odd dimensionalities $d=3$, $5$, $7$, and $9$, and to indices $k=2$ and $3$.

Figures \ref{fig:Error_vs_L}(a) and \ref{fig:Error_vs_L}(b) show the relative error $\left|\II_{L,d}^{(k)}[F(r)]/\II[F(r)]-1\right|$ versus $L$ for $\chi=2$ and $\chi=20$, respectively. Although not shown, in the case $\chi=20$ one can check that the error presents rapid oscillations as a function of $L$, except  for the combinations $(d,k)=(7,2)$, $(9,2)$, and $(9,3)$. To make cleaner the general picture, only integer values of $L$ are considered in Fig.\ \ref{fig:Error_vs_L}. We observe that an appropriate choice of $(d,k)$ can significantly reduce the error. In contrast to what is inferred from Ref.\ \cite{KV18}, the cases with $k=3$ generally perform better than with $k=2$. On the other hand, the optimal dimensionality $d$ depends on the correlation length: it is $d=3$ for $\chi=2$ and $d=7$ for $\chi=20$. Interestingly, when even values of $d$ are included, the best choices are $d=2$ (outperforming $d=3$) and $d=6$ (outperforming $d=7$) for $\chi=2$ and $\chi=20$, respectively (not shown).

To investigate the influence of the correlation length $\chi$ on the relative error, Figures \ref{fig:Error_vs_chi}(a) and \ref{fig:Error_vs_chi}(b) show the relative error $\left|\II_{L,d}^{(k)}[F(r)]/\II[F(r)]-1\right|$ versus $\chi$ for $L=5$ and $L=20$, respectively. The best behaviors are presented by $d=3$ if $L=5$ and by $d=7$ if $L=20$, in both cases with $k=3$. Although not shown, it turns out that $d=6$ outperforms $d=7$ if $L=20$.

As a second (and more realistic) illustrative example, let us take the exact solution of the Percus-Yevick (PY) integral equation for 3D hard spheres \cite{W63,T63,W64,AL66,S16,LNP_book_note_13_10}, which is exactly known for any packing fraction $\phi$. The results are displayed in Figs.\ \ref{fig:Error_vs_L_PY} and \ref{fig:Error_vs_eta_PY}. Again, the choices with $k=3$ are typically more accurate than with $k=2$. Also, as happened in the case of Eq.\ \eqref{h(r)}, the optimal choice of $d$ depends on the range of $h(r)$: while $d=3$ is appropriate for $\phi=0.2$, $d=7$ is preferable for $\phi=0.5$. When even dimensionalities are included (not shown), the best results are again obtained with $d=2$ and $d=6$ for $\phi=0.2$ and $\phi=0.5$, respectively.

\subsection{One-dimensional Kirkwood-Buff integrals}

In the case of more general functions $F(r)$ where the sought integral $\II[F(r)]$ is not known, the optimal choice of the embedding dimensionality $d$ and the index $k$ can be estimated by plotting $\II_{L,d}^{(k)}[F(r)]$ versus $L^{-3}$ for several combinations of $(d,k)$ and selecting the one with the smoothest variation allowing for an easy extrapolation to $L^{-3}\to 0$.

To illustrate this method, let us now consider the one-dimensional (1D) KB integral of hard rods (Tonks gas). In that case, $F(r)=2 h(r)$ is exactly known \cite{T36,SZK53,LZ71,HC04,S07,S16,FS17,LNP_book_note_13_08}, but we can pretend that the associated KB integral $\II[F(r)]$ is unknown. Figure \ref{fig:1D} shows the integrals $\II_{L,d}^{(k)}[F(r)]$ versus $L^{-3}$ at a packing fraction $\phi=0.8$. In all the cases, the integrals $\II_{L,d}^{(k)}[F(r)]$ are seen to converge to the exact value $\II[F(r)]=\phi-2=-1.2$. In general, the amplitudes of the oscillations are smaller with $k=2$ than with $k=3$ and decrease as the embedding dimensionality $d$ increases. On the other hand, the slopes of the lines around which the oscillations take place are smaller  with $k=3$ than with $k=2$ and decrease as $d$ decreases. Thus, the optimal choice of $(d,k)$ would depend on the accessible region  of $L$: if $L\sim 5$, $(d,k)=(9,3)$  seems to be a good choice for the extrapolation to $L^{-3}\to 0$, while $(d,k)=(7,3)$ seems preferable if $L\sim 10$.

\subsection{Three-dimensional  structure factors}
As shown by Eqs.\ \eqref{hq} and \eqref{Sq}, a relevant physical quantity directly related to integrals of the form \eqref{1} is the static structure factor of a liquid. In the case of 3D systems, $\widetilde{h}(q)=\II[F(r)]$ with $F(r)=(4\pi/q)r\sin (qr)h(r)$, which reduces to the KB integral in the limit $q\to 0$. If $q\neq 0$, the oscillations of $F(r)$ are not only due to $h(r)$ but also to the term $\sin(qr)$. Therefore, the optimization of the numerical or computational estimate of $\widetilde{h}(q)$ when $h(r)$ is known only for $r<L$ is again an extremely important goal.

Let us take once more the exact solution of the PY integral equation for hard spheres \cite{W63,T63,W64,AL66,S16,LNP_book_note_13_10} as a physically motivated benchmark to assess the performance of the approximations $\II[F(r)]\simeq \II_{L,d}^{(k)}[F(r)]$, this time as functions of the wave number $q$. Figure \ref{fig:Error_vs_q_PY} shows the relative error $\left|\II_{L,d}^{(k)}[F(r)]/\II[F(r)]-1\right|$ versus $q$, where $F(r)=(4\pi/q) r\sin(qr) h(r)$ and $h(r)$ is the PY pair correlation function  at a packing fraction $\phi=0.5$. The oscillations in the $q$-dependence of the relative error are typically smaller with $k=2$ than with $k=3$, and tend to smooth out as $d$ increases. As happened for the KB integrals (see Figs.\ \ref{fig:Error_vs_L}--\ref{fig:1D}), the error is generally smaller if $k=3$ than if $k=2$. As for the influence of the embedding dimensionality $d$, we see in Fig.\ \ref{fig:Error_vs_q_PY} that the best general estimates are obtained with $d=7$ and $d=3$ for a cutoff value $L=5$ and $L=10$, respectively, this time outperforming $d=6$ and $d=2$ (not shown).

\section{Conclusion}
\label{sec4}

In summary, the generalization to any embedding dimensionality $d$ and any index $k$ of the weight function $W_3^{(2)}(x)$, Eq.\ \eqref{3}, proposed in Ref.\ \cite{KV18} can significantly improve the cutoff estimate $\II_{L,d}^{(k)}[F(r)]$ of the improper integral $\II[F(r)]$ of an oscillatory function, even if $\II[F(r)]$ represents a KB integral corresponding to a 3D or 1D pair correlation function.

In the cases of KB integrals and structure factors, the results reported here show that an optimal choice of the index is $k=3$. As for the embedding dimensionality, its optimal value tends to increase as the correlation length increases, i.e., as the error due to the finite cutoff distance $L$ grows. As a practical compromise between simplicity and accuracy, a recommended weight function seems to be the one corresponding to $d=7$ and $k=3$, namely
\begin{align}
W_7^{(3)}(x)=&(1-x)^4\left(1+\frac{35x}{16}\right)\left(1+\frac{1225x^2}{256}\right)\nn
&\times\left(1+\frac{29x}{16}+\frac{5x^2}{4}+\frac{5x^3}{16}\right).
\end{align}

It must be noted that any method based on Eq.\ \eqref{2} with a weight function $0< W(x)<1$ ceases to be valid if the integrand $F(r)$ is not asymptotically an oscillatory function; if the magnitude of $F(r)$ decays monotonically, then the bare truncated integral \eqref{0} itself represents a better estimate than Eq.\ \eqref{2}. Thus, in the case of interaction potentials with an attractive tail, Eq.\ \eqref{2} must be discarded in the computational evaluation of KB integrals below the so-called Fisher-Widom line \cite{FW69,EHHPS93,VRL95,B96,DE00,TCV03,HRYS18}, where $h(r)$ decays monotonically. On the other hand, even in that case, Eq.\ \eqref{2} may be useful for the evaluation of the structure factor for moderate wave numbers.

Finally, let me point out that, while in this paper the addressed examples have been related to liquid state physics, given the ubiquitous appearance of integrals involving oscillatory integrands and semiinfinite intervals in many fields of physics, one would expect that the results  presented here will be useful for other physical problems as well.

\begin{acknowledgments}
The author is indebted to Mariano L\'opez de Haro for insightful discussions about the topic of this paper and to Arieh Iserles for calling Ref.\ \cite{DHI17} to his attention. Financial support from the Spanish Agencia Estatal de Investigaci\'on through Grant No.\ FIS2016-76359-P and the Junta de Extremadura
(Spain) through Grant No.\ GR18079, both partially financed by Fondo Europeo de Desarrollo Regional funds, is gratefully acknowledged.
\end{acknowledgments}

\bibliography{D:/Dropbox/Mis_Dropcumentos/bib_files/liquid}

\begin{thebibliography}{33}%
\makeatletter
\providecommand \@ifxundefined [1]{%
 \@ifx{#1\undefined}
}%
\providecommand \@ifnum [1]{%
 \ifnum #1\expandafter \@firstoftwo
 \else \expandafter \@secondoftwo
 \fi
}%
\providecommand \@ifx [1]{%
 \ifx #1\expandafter \@firstoftwo
 \else \expandafter \@secondoftwo
 \fi
}%
\providecommand \natexlab [1]{#1}%
\providecommand \enquote  [1]{``#1''}%
\providecommand \bibnamefont  [1]{#1}%
\providecommand \bibfnamefont [1]{#1}%
\providecommand \citenamefont [1]{#1}%
\providecommand \href@noop [0]{\@secondoftwo}%
\providecommand \href [0]{\begingroup \@sanitize@url \@href}%
\providecommand \@href[1]{\@@startlink{#1}\@@href}%
\providecommand \@@href[1]{\endgroup#1\@@endlink}%
\providecommand \@sanitize@url [0]{\catcode `\\12\catcode `\$12\catcode
  `\&12\catcode `\#12\catcode `\^12\catcode `\_12\catcode `\%12\relax}%
\providecommand \@@startlink[1]{}%
\providecommand \@@endlink[0]{}%
\providecommand \url  [0]{\begingroup\@sanitize@url \@url }%
\providecommand \@url [1]{\endgroup\@href {#1}{\urlprefix }}%
\providecommand \urlprefix  [0]{URL }%
\providecommand \Eprint [0]{\href }%
\providecommand \doibase [0]{http://dx.doi.org/}%
\providecommand \selectlanguage [0]{\@gobble}%
\providecommand \bibinfo  [0]{\@secondoftwo}%
\providecommand \bibfield  [0]{\@secondoftwo}%
\providecommand \translation [1]{[#1]}%
\providecommand \BibitemOpen [0]{}%
\providecommand \bibitemStop [0]{}%
\providecommand \bibitemNoStop [0]{.\EOS\space}%
\providecommand \EOS [0]{\spacefactor3000\relax}%
\providecommand \BibitemShut  [1]{\csname bibitem#1\endcsname}%
\let\auto@bib@innerbib\@empty
\bibitem [{\citenamefont {Kirkwood}\ and\ \citenamefont {Buff}(1951)}]{KB51}%
  \BibitemOpen
  \bibfield  {author} {\bibinfo {author} {\bibfnamefont {J.~G.}\ \bibnamefont
  {Kirkwood}}\ and\ \bibinfo {author} {\bibfnamefont {F.~P.}\ \bibnamefont
  {Buff}},\ }\bibfield  {title} {\enquote {\bibinfo {title} {The statistical
  mechanical theory of solutions. {I}},}\ }\href {\doibase 10.1063/1.1748352}
  {\bibfield  {journal} {\bibinfo  {journal} {J. Chem. Phys.}\ }\textbf
  {\bibinfo {volume} {19}},\ \bibinfo {pages} {774--777} (\bibinfo {year}
  {1951})}\BibitemShut {NoStop}%
\bibitem [{\citenamefont {Ben-Naim}(2006)}]{BN06}%
  \BibitemOpen
  \bibfield  {author} {\bibinfo {author} {\bibfnamefont {A.}~\bibnamefont
  {Ben-Naim}},\ }\href@noop {} {\emph {\bibinfo {title} {Molecular Theory of
  Solutions}}}\ (\bibinfo  {publisher} {Oxford University Press},\ \bibinfo
  {address} {Oxford, UK},\ \bibinfo {year} {2006})\BibitemShut {NoStop}%
\bibitem [{\citenamefont {Ben-Naim}\ and\ \citenamefont
  {Santos}(2009)}]{BNS09}%
  \BibitemOpen
  \bibfield  {author} {\bibinfo {author} {\bibfnamefont {A.}~\bibnamefont
  {Ben-Naim}}\ and\ \bibinfo {author} {\bibfnamefont {A.}~\bibnamefont
  {Santos}},\ }\bibfield  {title} {\enquote {\bibinfo {title} {Local and global
  properties of mixtures in one-dimensional systems. {II}. {Exact} results for
  the {Kirkwood}--{Buff} integrals},}\ }\href {\doibase 10.1063/1.3256234}
  {\bibfield  {journal} {\bibinfo  {journal} {J. Chem. Phys.}\ }\textbf
  {\bibinfo {volume} {131}},\ \bibinfo {pages} {{164}{512}} (\bibinfo {year}
  {2009})}\BibitemShut {NoStop}%
\bibitem [{\citenamefont {Barker}\ and\ \citenamefont
  {Henderson}(1976)}]{BH76}%
  \BibitemOpen
  \bibfield  {author} {\bibinfo {author} {\bibfnamefont {J.~A.}\ \bibnamefont
  {Barker}}\ and\ \bibinfo {author} {\bibfnamefont {D.}~\bibnamefont
  {Henderson}},\ }\bibfield  {title} {\enquote {\bibinfo {title} {What is
  ``liquid''? {Understanding} the states of matter},}\ }\href {\doibase
  10.1103/RevModPhys.48.587} {\bibfield  {journal} {\bibinfo  {journal} {Rev.
  Mod. Phys.}\ }\textbf {\bibinfo {volume} {48}},\ \bibinfo {pages} {587--671}
  (\bibinfo {year} {1976})}\BibitemShut {NoStop}%
\bibitem [{\citenamefont {Hansen}\ and\ \citenamefont {McDonald}(2006)}]{HM06}%
  \BibitemOpen
  \bibfield  {author} {\bibinfo {author} {\bibfnamefont {J.-P.}\ \bibnamefont
  {Hansen}}\ and\ \bibinfo {author} {\bibfnamefont {I.~R.}\ \bibnamefont
  {McDonald}},\ }\href@noop {} {\emph {\bibinfo {title} {{Theory of Simple
  Liquids}}}},\ \bibinfo {edition} {3rd}\ ed.\ (\bibinfo  {publisher}
  {Academic},\ \bibinfo {address} {London},\ \bibinfo {year}
  {2006})\BibitemShut {NoStop}%
\bibitem [{\citenamefont {Santos}(2016)}]{S16}%
  \BibitemOpen
  \bibfield  {author} {\bibinfo {author} {\bibfnamefont {A.}~\bibnamefont
  {Santos}},\ }\href@noop {} {\emph {\bibinfo {title} {{A Concise Course on the
  Theory of Classical Liquids. Basics and Selected Topics}}}},\ \bibinfo
  {series} {Lecture Notes in Physics}, Vol.\ \bibinfo {volume} {923}\ (\bibinfo
   {publisher} {Springer},\ \bibinfo {address} {New York},\ \bibinfo {year}
  {2016})\BibitemShut {NoStop}%
\bibitem [{\citenamefont {Dea{\~n}o}\ \emph {et~al.}(2017)\citenamefont
  {Dea{\~n}o}, \citenamefont {Huybrechs},\ and\ \citenamefont
  {Iserles}}]{DHI17}%
  \BibitemOpen
  \bibfield  {author} {\bibinfo {author} {\bibfnamefont {A.}~\bibnamefont
  {Dea{\~n}o}}, \bibinfo {author} {\bibfnamefont {D.}~\bibnamefont
  {Huybrechs}}, \ and\ \bibinfo {author} {\bibfnamefont {A.}~\bibnamefont
  {Iserles}},\ }\href@noop {} {\emph {\bibinfo {title} {Computing Highly
  Oscillatory Integrals}}}\ (\bibinfo  {publisher} {Society for Industrial and
  Applied Mathematics},\ \bibinfo {address} {Philadelphia, PA},\ \bibinfo
  {year} {2017})\BibitemShut {NoStop}%
\bibitem [{\citenamefont {Kr\"{u}ger}\ \emph {et~al.}(2013)\citenamefont
  {Kr\"{u}ger}, \citenamefont {Schnell}, \citenamefont {Bedeaux}, \citenamefont
  {Kjelstrup}, \citenamefont {Vlugt},\ and\ \citenamefont {Simon}}]{KSBKVS13}%
  \BibitemOpen
  \bibfield  {author} {\bibinfo {author} {\bibfnamefont {P.}~\bibnamefont
  {Kr\"{u}ger}}, \bibinfo {author} {\bibfnamefont {S.~K.}\ \bibnamefont
  {Schnell}}, \bibinfo {author} {\bibfnamefont {D.}~\bibnamefont {Bedeaux}},
  \bibinfo {author} {\bibfnamefont {S.}~\bibnamefont {Kjelstrup}}, \bibinfo
  {author} {\bibfnamefont {T.~J.~H.}\ \bibnamefont {Vlugt}}, \ and\ \bibinfo
  {author} {\bibfnamefont {J.-M.}\ \bibnamefont {Simon}},\ }\bibfield  {title}
  {\enquote {\bibinfo {title} {Kirkwood--{B}uff integrals for finite
  volumes},}\ }\href {\doibase 10.1021/jz301992u} {\bibfield  {journal}
  {\bibinfo  {journal} {J. Phys. Chem. Lett.}\ }\textbf {\bibinfo {volume}
  {4}},\ \bibinfo {pages} {235--238} (\bibinfo {year} {2013})}\BibitemShut
  {NoStop}%
\bibitem [{\citenamefont {Kr\"{u}ger}\ and\ \citenamefont
  {Vlugt}(2018)}]{KV18}%
  \BibitemOpen
  \bibfield  {author} {\bibinfo {author} {\bibfnamefont {P.}~\bibnamefont
  {Kr\"{u}ger}}\ and\ \bibinfo {author} {\bibfnamefont {T.~J.~H.}\ \bibnamefont
  {Vlugt}},\ }\bibfield  {title} {\enquote {\bibinfo {title} {Size and shape
  dependence of finite-volume {K}irkwood-{B}uff integrals},}\ }\href {\doibase
  10.1103/PhysRevE.97.051301} {\bibfield  {journal} {\bibinfo  {journal} {Phys.
  Rev. E}\ }\textbf {\bibinfo {volume} {97}},\ \bibinfo {pages} {051301(R)}
  (\bibinfo {year} {2018})}\BibitemShut {NoStop}%
\bibitem [{\citenamefont {Abramowitz}\ and\ \citenamefont
  {Stegun}(1972)}]{AS72}%
  \BibitemOpen
  \bibinfo {editor} {\bibfnamefont {M.}~\bibnamefont {Abramowitz}}\ and\
  \bibinfo {editor} {\bibfnamefont {I.~A.}\ \bibnamefont {Stegun}},\ eds.,\
  \href@noop {} {\emph {\bibinfo {title} {{Handbook of Mathematical
  Functions}}}}\ (\bibinfo  {publisher} {Dover},\ \bibinfo {address} {New
  York},\ \bibinfo {year} {1972})\BibitemShut {NoStop}%
\bibitem [{\citenamefont {Olver}\ \emph {et~al.}(2010)\citenamefont {Olver},
  \citenamefont {Lozier}, \citenamefont {Boisvert},\ and\ \citenamefont
  {Clark}}]{OLBC10}%
  \BibitemOpen
  \bibinfo {editor} {\bibfnamefont {F.~W.~J.}\ \bibnamefont {Olver}}, \bibinfo
  {editor} {\bibfnamefont {D.~W.}\ \bibnamefont {Lozier}}, \bibinfo {editor}
  {\bibfnamefont {R.~F.}\ \bibnamefont {Boisvert}}, \ and\ \bibinfo {editor}
  {\bibfnamefont {C.~W.}\ \bibnamefont {Clark}},\ eds.,\ \href@noop {} {\emph
  {\bibinfo {title} {NIST Handbook of Mathematical Functions}}}\ (\bibinfo
  {publisher} {Cambridge University Press},\ \bibinfo {address} {New York},\
  \bibinfo {year} {2010})\BibitemShut {NoStop}%
\bibitem [{\citenamefont {Baus}\ and\ \citenamefont {Colot}(1987)}]{BC87}%
  \BibitemOpen
  \bibfield  {author} {\bibinfo {author} {\bibfnamefont {M.}~\bibnamefont
  {Baus}}\ and\ \bibinfo {author} {\bibfnamefont {J.~L.}\ \bibnamefont
  {Colot}},\ }\bibfield  {title} {\enquote {\bibinfo {title} {{Thermodynamics
  and structure of a fluid of hard rods, disks, spheres, or hyperspheres from
  rescaled virial expansions}},}\ }\href {\doibase 10.1103/PhysRevA.36.3912}
  {\bibfield  {journal} {\bibinfo  {journal} {Phys. Rev. A}\ }\textbf {\bibinfo
  {volume} {36}},\ \bibinfo {pages} {3912--3925} (\bibinfo {year}
  {1987})}\BibitemShut {NoStop}%
\bibitem [{\citenamefont {Torquato}\ and\ \citenamefont
  {Stillinger}(2003)}]{TS03}%
  \BibitemOpen
  \bibfield  {author} {\bibinfo {author} {\bibfnamefont {S.}~\bibnamefont
  {Torquato}}\ and\ \bibinfo {author} {\bibfnamefont {F.~H.}\ \bibnamefont
  {Stillinger}},\ }\bibfield  {title} {\enquote {\bibinfo {title} {Local
  density fluctuations, hyperuniformity, and order metrics},}\ }\href {\doibase
  10.1103/PhysRevE.68.041113} {\bibfield  {journal} {\bibinfo  {journal} {Phys.
  Rev. E}\ }\textbf {\bibinfo {volume} {68}},\ \bibinfo {pages} {{041}{113}}
  (\bibinfo {year} {2003})}\BibitemShut {NoStop}%
\bibitem [{\citenamefont {Torquato}\ and\ \citenamefont
  {Stillinger}(2006)}]{TS06}%
  \BibitemOpen
  \bibfield  {author} {\bibinfo {author} {\bibfnamefont {S.}~\bibnamefont
  {Torquato}}\ and\ \bibinfo {author} {\bibfnamefont {F.~H.}\ \bibnamefont
  {Stillinger}},\ }\bibfield  {title} {\enquote {\bibinfo {title} {New
  conjectural lower bounds on the optimal density of sphere packings},}\ }\href
  {\doibase 10.1080/10586458.2006.10128964} {\bibfield  {journal} {\bibinfo
  {journal} {Exp. Math.}\ }\textbf {\bibinfo {volume} {15}},\ \bibinfo {pages}
  {307--331} (\bibinfo {year} {2006})}\BibitemShut {NoStop}%
\bibitem [{\citenamefont {Wertheim}(1963)}]{W63}%
  \BibitemOpen
  \bibfield  {author} {\bibinfo {author} {\bibfnamefont {M.~S.}\ \bibnamefont
  {Wertheim}},\ }\bibfield  {title} {\enquote {\bibinfo {title} {Exact solution
  of the {Percus--Yevick} integral equation for hard spheres},}\ }\href
  {\doibase 10.1103/PhysRevLett.10.321.} {\bibfield  {journal} {\bibinfo
  {journal} {Phys. Rev. Lett.}\ }\textbf {\bibinfo {volume} {10}},\ \bibinfo
  {pages} {321--323} (\bibinfo {year} {1963})}\BibitemShut {NoStop}%
\bibitem [{\citenamefont {Thiele}(1963)}]{T63}%
  \BibitemOpen
  \bibfield  {author} {\bibinfo {author} {\bibfnamefont {E.}~\bibnamefont
  {Thiele}},\ }\bibfield  {title} {\enquote {\bibinfo {title} {Equation of
  state for hard spheres},}\ }\href {\doibase 10.1063/1.1734272} {\bibfield
  {journal} {\bibinfo  {journal} {J. Chem. Phys.}\ }\textbf {\bibinfo {volume}
  {39}},\ \bibinfo {pages} {474--479} (\bibinfo {year} {1963})}\BibitemShut
  {NoStop}%
\bibitem [{\citenamefont {Wertheim}(1964)}]{W64}%
  \BibitemOpen
  \bibfield  {author} {\bibinfo {author} {\bibfnamefont {M.~S.}\ \bibnamefont
  {Wertheim}},\ }\bibfield  {title} {\enquote {\bibinfo {title} {Analytic
  solution of the {Percus--Yevick} equation},}\ }\href {\doibase
  10.1063/1.1704158} {\bibfield  {journal} {\bibinfo  {journal} {J. Math.
  Phys.}\ }\textbf {\bibinfo {volume} {5}},\ \bibinfo {pages} {643--651}
  (\bibinfo {year} {1964})}\BibitemShut {NoStop}%
\bibitem [{\citenamefont {Ashcroft}\ and\ \citenamefont {Lekner}(1966)}]{AL66}%
  \BibitemOpen
  \bibfield  {author} {\bibinfo {author} {\bibfnamefont {N.~W.}\ \bibnamefont
  {Ashcroft}}\ and\ \bibinfo {author} {\bibfnamefont {J.}~\bibnamefont
  {Lekner}},\ }\bibfield  {title} {\enquote {\bibinfo {title} {Structure and
  resistivity of liquid metals},}\ }\href {\doibase 10.1103/PhysRev.145.83}
  {\bibfield  {journal} {\bibinfo  {journal} {Phys. Rev.}\ }\textbf {\bibinfo
  {volume} {145}},\ \bibinfo {pages} {83--90} (\bibinfo {year}
  {1966})}\BibitemShut {NoStop}%
\bibitem [{\citenamefont {Santos}(2013)}]{LNP_book_note_13_10}%
  \BibitemOpen
  \bibfield  {author} {\bibinfo {author} {\bibfnamefont {A.}~\bibnamefont
  {Santos}},\ }\href@noop {} {} (\bibinfo {year} {2013}),\ \bibinfo {note}
  {``Radial Distribution Function for Hard Spheres'', Wolfram Demonstrations
  Project,
  \begin{scriptsize}\url{http://demonstrations.wolfram.com/RadialDistributionFunctionForHardSpheres/}\end{scriptsize}}\BibitemShut
  {NoStop}%
\bibitem [{\citenamefont {Tonks}(1936)}]{T36}%
  \BibitemOpen
  \bibfield  {author} {\bibinfo {author} {\bibfnamefont {L.}~\bibnamefont
  {Tonks}},\ }\bibfield  {title} {\enquote {\bibinfo {title} {The complete
  equation of state of one, two and three-dimensional gases of hard elastic
  spheres},}\ }\href {\doibase 10.1103/PhysRev.50.955} {\bibfield  {journal}
  {\bibinfo  {journal} {Phys. Rev.}\ }\textbf {\bibinfo {volume} {50}},\
  \bibinfo {pages} {955--963} (\bibinfo {year} {1936})}\BibitemShut {NoStop}%
\bibitem [{\citenamefont {Salsburg}\ \emph {et~al.}(1953)\citenamefont
  {Salsburg}, \citenamefont {Zwanzig},\ and\ \citenamefont {Kirkwood}}]{SZK53}%
  \BibitemOpen
  \bibfield  {author} {\bibinfo {author} {\bibfnamefont {Z.~W.}\ \bibnamefont
  {Salsburg}}, \bibinfo {author} {\bibfnamefont {R.~W.}\ \bibnamefont
  {Zwanzig}}, \ and\ \bibinfo {author} {\bibfnamefont {J.~G.}\ \bibnamefont
  {Kirkwood}},\ }\bibfield  {title} {\enquote {\bibinfo {title} {Molecular
  distribution functions in a one-dimensional fluid},}\ }\href {\doibase
  10.1063/1.1699116} {\bibfield  {journal} {\bibinfo  {journal} {J. Chem.
  Phys.}\ }\textbf {\bibinfo {volume} {21}},\ \bibinfo {pages} {1098--1107}
  (\bibinfo {year} {1953})}\BibitemShut {NoStop}%
\bibitem [{\citenamefont {Lebowitz}\ and\ \citenamefont {Zomick}(1971)}]{LZ71}%
  \BibitemOpen
  \bibfield  {author} {\bibinfo {author} {\bibfnamefont {J.~L.}\ \bibnamefont
  {Lebowitz}}\ and\ \bibinfo {author} {\bibfnamefont {D.}~\bibnamefont
  {Zomick}},\ }\bibfield  {title} {\enquote {\bibinfo {title} {Mixtures of hard
  spheres with nonadditive diameters: {S}ome exact results and solution of {PY}
  equation},}\ }\href {\doibase 10.1063/1.1675348} {\bibfield  {journal}
  {\bibinfo  {journal} {J. Chem. Phys.}\ }\textbf {\bibinfo {volume} {54}},\
  \bibinfo {pages} {3335--3346} (\bibinfo {year} {1971})}\BibitemShut {NoStop}%
\bibitem [{\citenamefont {Heying}\ and\ \citenamefont {Corti}(2004)}]{HC04}%
  \BibitemOpen
  \bibfield  {author} {\bibinfo {author} {\bibfnamefont {M.}~\bibnamefont
  {Heying}}\ and\ \bibinfo {author} {\bibfnamefont {D.~S.}\ \bibnamefont
  {Corti}},\ }\bibfield  {title} {\enquote {\bibinfo {title} {The
  one-dimensional fully non-additive binary hard rod mixture: exact
  thermophysical properties},}\ }\href {\doibase 10.1016/j.fluid.2004.02.018}
  {\bibfield  {journal} {\bibinfo  {journal} {Fluid Phase Equil.}\ }\textbf
  {\bibinfo {volume} {220}},\ \bibinfo {pages} {85--103} (\bibinfo {year}
  {2004})}\BibitemShut {NoStop}%
\bibitem [{\citenamefont {Santos}(2007)}]{S07}%
  \BibitemOpen
  \bibfield  {author} {\bibinfo {author} {\bibfnamefont {A.}~\bibnamefont
  {Santos}},\ }\bibfield  {title} {\enquote {\bibinfo {title} {Exact bulk
  correlation functions in one-dimensional nonadditive hard-core mixtures},}\
  }\href {\doibase 10.1103/PhysRevE.76.062201} {\bibfield  {journal} {\bibinfo
  {journal} {Phys. Rev. E}\ }\textbf {\bibinfo {volume} {76}},\ \bibinfo
  {pages} {{062}{201}} (\bibinfo {year} {2007})}\BibitemShut {NoStop}%
\bibitem [{\citenamefont {Fantoni}\ and\ \citenamefont {Santos}(2017)}]{FS17}%
  \BibitemOpen
  \bibfield  {author} {\bibinfo {author} {\bibfnamefont {R.}~\bibnamefont
  {Fantoni}}\ and\ \bibinfo {author} {\bibfnamefont {A.}~\bibnamefont
  {Santos}},\ }\bibfield  {title} {\enquote {\bibinfo {title} {One-dimensional
  fluids with second nearest-neighbor interactions},}\ }\href {\doibase
  10.1007/s10955-017-1908-6} {\bibfield  {journal} {\bibinfo  {journal} {J.
  Stat. Phys.}\ }\textbf {\bibinfo {volume} {169}},\ \bibinfo {pages}
  {1171--1201} (\bibinfo {year} {2017})}\BibitemShut {NoStop}%
\bibitem [{\citenamefont {Santos}(2012)}]{LNP_book_note_13_08}%
  \BibitemOpen
  \bibfield  {author} {\bibinfo {author} {\bibfnamefont {A.}~\bibnamefont
  {Santos}},\ }\href@noop {} {} (\bibinfo {year} {2012}),\ \bibinfo {note}
  {``Radial Distribution Function for Sticky Hard Rods'', Wolfram
  Demonstrations Project,
  \begin{scriptsize}\url{http://demonstrations.wolfram.com/RadialDistributionFunctionForStickyHardRods/}\end{scriptsize}}\BibitemShut
  {NoStop}%
\bibitem [{\citenamefont {Fisher}\ and\ \citenamefont {Widom}(1969)}]{FW69}%
  \BibitemOpen
  \bibfield  {author} {\bibinfo {author} {\bibfnamefont {M.~E.}\ \bibnamefont
  {Fisher}}\ and\ \bibinfo {author} {\bibfnamefont {B.}~\bibnamefont {Widom}},\
  }\bibfield  {title} {\enquote {\bibinfo {title} {Decay of correlations in
  linear systems},}\ }\href {\doibase 10.1063/1.1671624} {\bibfield  {journal}
  {\bibinfo  {journal} {J. Chem. Phys.}\ }\textbf {\bibinfo {volume} {50}},\
  \bibinfo {pages} {3756--3772} (\bibinfo {year} {1969})}\BibitemShut {NoStop}%
\bibitem [{\citenamefont {Evans}\ \emph {et~al.}(1993)\citenamefont {Evans},
  \citenamefont {Henderson}, \citenamefont {Hoyle}, \citenamefont {Parry},\
  and\ \citenamefont {Sabeur}}]{EHHPS93}%
  \BibitemOpen
  \bibfield  {author} {\bibinfo {author} {\bibfnamefont {R.}~\bibnamefont
  {Evans}}, \bibinfo {author} {\bibfnamefont {J.~R.}\ \bibnamefont
  {Henderson}}, \bibinfo {author} {\bibfnamefont {D.~C.}\ \bibnamefont
  {Hoyle}}, \bibinfo {author} {\bibfnamefont {A.~O.}\ \bibnamefont {Parry}}, \
  and\ \bibinfo {author} {\bibfnamefont {Z.~A.}\ \bibnamefont {Sabeur}},\
  }\bibfield  {title} {\enquote {\bibinfo {title} {Asymptotic decay of liquid
  structure: oscillatory liquid-vapour density profiles and the
  {F}isher--{W}idom line},}\ }\href {\doibase 10.1080/00268979300102621}
  {\bibfield  {journal} {\bibinfo  {journal} {Mol. Phys.}\ }\textbf {\bibinfo
  {volume} {80}},\ \bibinfo {pages} {755--775} (\bibinfo {year}
  {1993})}\BibitemShut {NoStop}%
\bibitem [{\citenamefont {Vega}\ \emph {et~al.}(1995)\citenamefont {Vega},
  \citenamefont {Rull},\ and\ \citenamefont {Lago}}]{VRL95}%
  \BibitemOpen
  \bibfield  {author} {\bibinfo {author} {\bibfnamefont {C.}~\bibnamefont
  {Vega}}, \bibinfo {author} {\bibfnamefont {L.~F.}\ \bibnamefont {Rull}}, \
  and\ \bibinfo {author} {\bibfnamefont {S.}~\bibnamefont {Lago}},\ }\bibfield
  {title} {\enquote {\bibinfo {title} {Location of the {F}isher-{W}idom line
  for systems interacting through short-ranged potentials},}\ }\href {\doibase
  10.1103/PhysRevE.51.3146} {\bibfield  {journal} {\bibinfo  {journal} {Phys.
  Rev. E}\ }\textbf {\bibinfo {volume} {51}},\ \bibinfo {pages} {3146--3155}
  (\bibinfo {year} {1995})}\BibitemShut {NoStop}%
\bibitem [{\citenamefont {Brown}(1996)}]{B96}%
  \BibitemOpen
  \bibfield  {author} {\bibinfo {author} {\bibfnamefont {W.~E.}\ \bibnamefont
  {Brown}},\ }\bibfield  {title} {\enquote {\bibinfo {title} {The
  {F}isher--{W}idom line for a hard core attractive {Y}ukawa fluid},}\ }\href
  {\doibase 10.1080/00268979650026541} {\bibfield  {journal} {\bibinfo
  {journal} {Mol. Phys.}\ }\textbf {\bibinfo {volume} {88}},\ \bibinfo {pages}
  {579--584} (\bibinfo {year} {1996})}\BibitemShut {NoStop}%
\bibitem [{\citenamefont {Dijkstra}\ and\ \citenamefont {Evans}(2000)}]{DE00}%
  \BibitemOpen
  \bibfield  {author} {\bibinfo {author} {\bibfnamefont {M.}~\bibnamefont
  {Dijkstra}}\ and\ \bibinfo {author} {\bibfnamefont {R.}~\bibnamefont
  {Evans}},\ }\bibfield  {title} {\enquote {\bibinfo {title} {A simulation
  study of the decay of the pair correlation function in simple fluids},}\
  }\href {\doibase 0.1063/1.480598} {\bibfield  {journal} {\bibinfo  {journal}
  {J. Chem. Phys.}\ }\textbf {\bibinfo {volume} {112}},\ \bibinfo {pages}
  {1449--1456} (\bibinfo {year} {2000})}\BibitemShut {NoStop}%
\bibitem [{\citenamefont {Tarazona}\ \emph {et~al.}(2003)\citenamefont
  {Tarazona}, \citenamefont {Chac\'on},\ and\ \citenamefont {Velasco}}]{TCV03}%
  \BibitemOpen
  \bibfield  {author} {\bibinfo {author} {\bibfnamefont {P.}~\bibnamefont
  {Tarazona}}, \bibinfo {author} {\bibfnamefont {E.}~\bibnamefont {Chac\'on}},
  \ and\ \bibinfo {author} {\bibfnamefont {E.}~\bibnamefont {Velasco}},\
  }\bibfield  {title} {\enquote {\bibinfo {title} {The {F}isher--{W}idom line
  for systems with low melting temperature},}\ }\href {\doibase
  10.1080/0026897031000068550} {\bibfield  {journal} {\bibinfo  {journal} {Mol.
  Phys.}\ }\textbf {\bibinfo {volume} {101}},\ \bibinfo {pages} {1595--1603}
  (\bibinfo {year} {2003})}\BibitemShut {NoStop}%
\bibitem [{\citenamefont {{L\'opez de Haro}}\ \emph {et~al.}(2018)\citenamefont
  {{L\'opez de Haro}}, \citenamefont {Rodr\'{\i}guez-Rivas}, \citenamefont
  {Yuste},\ and\ \citenamefont {Santos}}]{HRYS18}%
  \BibitemOpen
  \bibfield  {author} {\bibinfo {author} {\bibfnamefont {M.}~\bibnamefont
  {{L\'opez de Haro}}}, \bibinfo {author} {\bibfnamefont {A.}~\bibnamefont
  {Rodr\'{\i}guez-Rivas}}, \bibinfo {author} {\bibfnamefont {S.~B.}\
  \bibnamefont {Yuste}}, \ and\ \bibinfo {author} {\bibfnamefont
  {A.}~\bibnamefont {Santos}},\ }\bibfield  {title} {\enquote {\bibinfo {title}
  {Structural properties of the {J}agla fluid},}\ }\href {\doibase
  10.1103/PhysRevE.98.012138} {\bibfield  {journal} {\bibinfo  {journal} {Phys.
  Rev. E}\ }\textbf {\bibinfo {volume} {98}},\ \bibinfo {pages} {012138}
  (\bibinfo {year} {2018})}\BibitemShut {NoStop}%
\end{thebibliography}%

\end{document}